\documentclass[aps,prl,reprint,superscriptaddress]{revtex4-2}

\usepackage{graphicx}

\begin{document}

\title{Robust Fermi-Surface Morphology of CeRhIn$_5$ across the Putative Field-Induced Quantum Critical Point}

\author{S.~Mishra}
\affiliation{Laboratoire National des Champs Magn\'{e}tiques Intenses (LNCMI-EMFL), CNRS, UGA, 38042 Grenoble, France}

\author{J.~Hornung}
\affiliation{Hochfeld-Magnetlabor Dresden (HLD-EMFL) and W\"{u}rzburg-Dresden Cluster of Excellence ct.qmat, Helmholtz-Zentrum Dresden-Rossendorf, 01328 Dresden, Germany}
\affiliation{Institut f\"ur Festk\"orper- und Materialphysik, TU Dresden, 01062 Dresden, Germany}

\author{M.~Raba}
\altaffiliation[Present address: ]{IRIG/DEPHY/MEM-MDN, CEA-Grenoble, 17 Avenue des Martyrs, 38000 Grenoble, France}
\affiliation{Laboratoire National des Champs Magn\'{e}tiques Intenses (LNCMI-EMFL), CNRS, UGA, 38042 Grenoble, France}

\author{J.~Klotz}
\affiliation{Hochfeld-Magnetlabor Dresden (HLD-EMFL) and W\"{u}rzburg-Dresden Cluster of Excellence ct.qmat, Helmholtz-Zentrum Dresden-Rossendorf, 01328 Dresden, Germany}

\author{T.~F\"{o}rster}
\affiliation{Hochfeld-Magnetlabor Dresden (HLD-EMFL) and W\"{u}rzburg-Dresden Cluster of Excellence ct.qmat, Helmholtz-Zentrum Dresden-Rossendorf, 01328 Dresden, Germany}

\author{H.~Harima}
\affiliation{Graduate School of Science, Kobe University, Kobe 657-8501, Japan}

\author{D. Aoki}
\affiliation{Institute for Materials Research, Tohoku University, Oarai, Ibaraki, 311-1313, Japan}

\author{J.~Wosnitza}
\affiliation{Hochfeld-Magnetlabor Dresden (HLD-EMFL) and W\"{u}rzburg-Dresden Cluster of Excellence ct.qmat, Helmholtz-Zentrum Dresden-Rossendorf, 01328 Dresden, Germany}
\affiliation{Institut f\"ur Festk\"orper- und Materialphysik, TU Dresden, 01062 Dresden, Germany}

\author{A.~McCollam}
\affiliation{High Field Magnet Laboratory (HFML-EMFL), Radboud University, 6525 ED Nijmegen, The Netherlands}

\author{I.~Sheikin}
\email[]{ilya.sheikin@lncmi.cnrs.fr}
\affiliation{Laboratoire National des Champs Magn\'{e}tiques Intenses (LNCMI-EMFL), CNRS, UGA, 38042 Grenoble, France}

\date{\today}

\begin{abstract}
We report a comprehensive de Haas--van Alphen (dHvA) study of the heavy-fermion material CeRhIn$_5$ in magnetic fields up to 70~T. Several dHvA frequencies gradually emerge at high fields as a result of magnetic breakdown. Among them is the thermodynamically important $\beta_1$ branch, which has not been observed so far. Comparison of our angule-dependent dHvA spectra with those of the non-$4f$ compound LaRhIn$_5$ and with band-structure calculations evidences that the Ce $4f$ electrons in CeRhIn$_5$ remain localized over the whole field range. This rules out any significant Fermi-surface reconstruction, either at the suggested nematic phase transition at $B^{*}\approx$ 30~T or at the putative quantum critical point at $B_c \simeq$ 50~T. Our results rather demonstrate the robustness of the Fermi surface and the localized nature of the 4$f$ electrons inside and outside of the antiferromagnetic phase.
\end{abstract}

\maketitle

Rare-earth-based materials are now widely recognized as an ideal playground for exploration of the fascinating physics that develops around a quantum critical point (QCP), a second-order phase transition at zero temperature~\cite{Gegenwart2008,Si2001,Senthil2004}. In Ce-based compounds, such a QCP can be
induced by hydrostatic pressure, chemical doping, or magnetic field, where it typically separates a magnetically ordered state from a nonmagnetic ground state. In spite of numerous experimental investigations of such systems in the vicinity of a QCP, the details of what drives the QCP remain the subject of much theoretical debate. There are currently two fundamentally different theoretical models that attempt to describe the physics of antiferromagnetic (AFM) QCPs in heavy-fermion (HF) materials. The models can be distinguished by whether the Ce 4$f$ electrons are localized or itinerant on the either side of a QCP. Here, ``itinerant" means that the $f$ electrons are fully hybridized with the conduction electrons and therefore contribute to the Fermi surface (FS). The first type of QCP, referred to as a spin-density-wave QCP~\cite{Hertz1976,Millis1993}, assumes the $f$ electrons to be itinerant on both sides of a QCP. In this case, if delocalization of $f$ electrons occurs,  it occurs inside the magnetic phase. The second type of QCP, known as a Kondo-breakdown QCP~\cite{Si2001,Paschen2004,Coleman2001,Gegenwart2007,Friedemann2010}, suggests that a transition from itinerant to localized 4$f$ electrons occurs precisely at the QCP. Furthermore, the effective masses of the conduction quasiparticles are expected to diverge upon approaching this type of QCP.

Since the FSs with itinerant and localized $f$ electrons possess a different size and morphology, the two cases can be easily distinguished experimentally by performing quantum-oscillation measurements such as the de Haas--van Alphen (dHvA) effect. A comparison of the experimental angule-dependent dHvA spectra with results of band-structure calculations, both for localized and itinerant $4f$ electrons, allows us to distinguish between both scenarios. For Ce-based compounds, a comparison can also be made using experimental results obtained on La-based analogs, which serve as $f$-localized references, since the electronic structures of Ce and La differ by only one $f$ electron.

CeRhIn$_5$ is one of the best-studied HF materials. This tetragonal AFM compound with $T_N =$ 3.8~K can be tuned to a QCP by pressure~\cite{Hegger2000,Mito2003,Knebel2006,Park2006}, chemical substitution~\cite{Pagliuso2001,Zapf2001,Bauer2006}, and magnetic field~\cite{Jiao2015,Jiao2019}. The pressure-induced QCP in CeRhIn$_5$ is now considered to be a textbook example of the Kondo-breakdown type. Several dHvA experiments evidence that the $f$ electrons of CeRhIn$_5$ are localized at ambient pressure~\cite{Shishido2002,Harrison2004}, although some of the theoretically predicted dHvA frequencies were not experimentally observed~\cite{Shishido2002}. As the critical pressure for the suppression of antiferromagnetism, $P_c =$ 2.3~GPa, is reached, all dHvA frequencies observed at $P < P_c$ change discontinuously, signaling an abrupt FS reconstruction as a consequence of the $f$-electron delocalization~\cite{Shishido2005}. In addition, the effective masses diverge at $P_c$, further supporting the Kondo-breakdown scenario. A similar discontinuous change of the dHvA frequencies was observed upon substituting Rh by Co in CeRh$_{1-x}$Co$_x$In$_5$~\cite{Goh2008}. However, the FS reconstruction does not occur at the critical concentration $x_c \approx$ 0.8, where the AFM order is suppressed, but deep inside the AFM state at $x \simeq$ 0.4, where the AFM order alters its character and superconductivity emerges to coexist with antiferromagnetism. Furthermore, the effective masses do not diverge here. Thus, the substitution-induced QCP appears to be of the spin-density-wave type.

\begin{figure}[h!]
\includegraphics[width=\columnwidth]{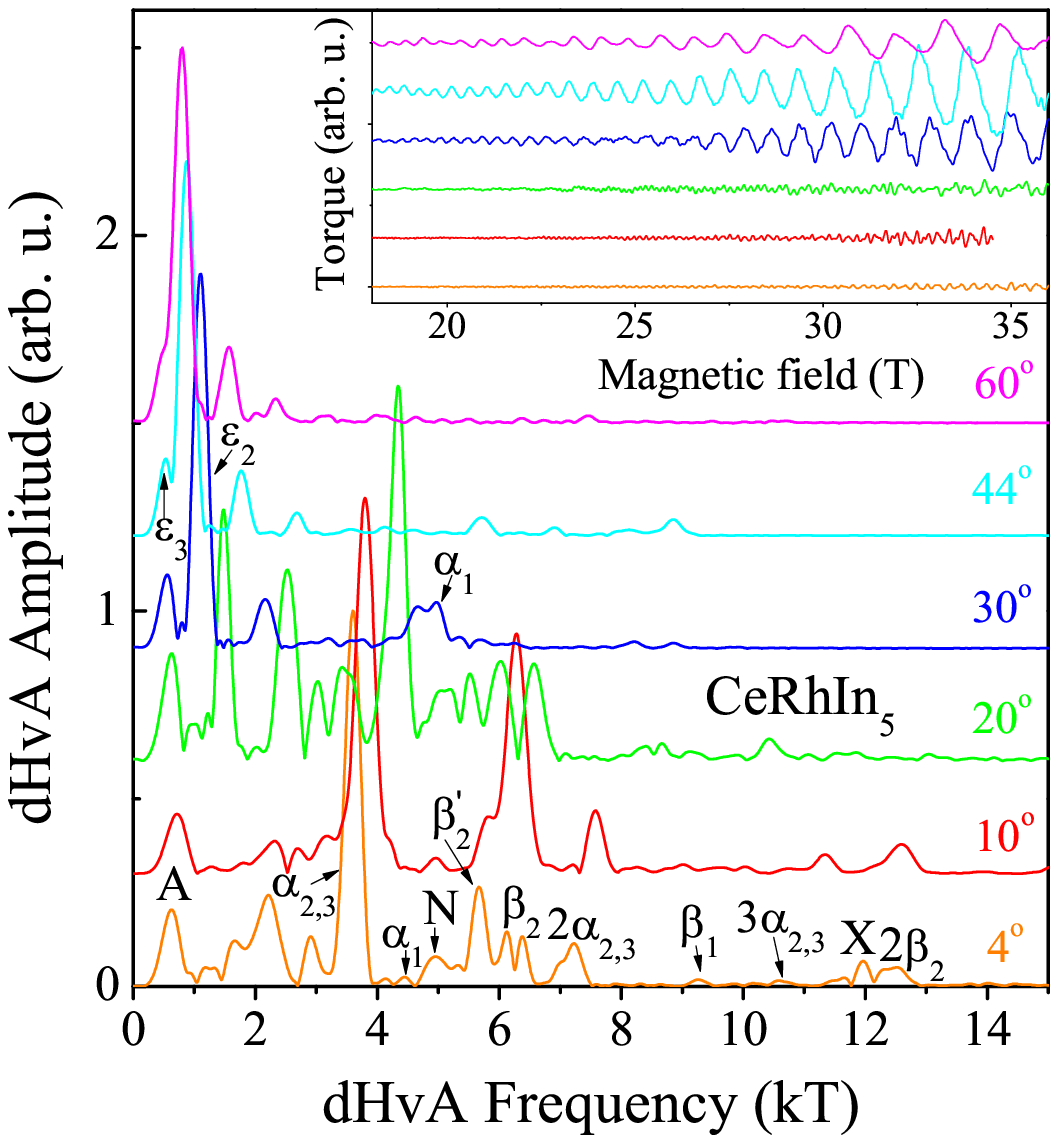}
\caption{\label{fig:ADofQOFFT} Fourier spectra of the dHvA oscillations (shown in the inset) of CeRhIn$_5$ for a magnetic field applied at various angles, $\theta$, from the $\textit{c}$ toward the $a$ axis at $T =$ 30~mK. The curves are shifted vertically for clarity. The FFTs are performed over the field interval from $B_{min}$ = 29~T to $B_{max}$ = 36~T, except at 10$^\circ$, where it is 29 - 34.5~T (see the inset) due to experimental constraints. All FFTs are normalized to the strongest dHvA spectral peak. An equivalent figure showing the data and FFTs for LaRhIn$_5$ is shown in the Supplemental Material~\cite{Suppl}.}

\end{figure}

\nocite{Gunnarsson1976,Haga,Thompson2001,Moshopoulou2001,Moshopoulou2002,Koelling1977,Shoenberg2009,Joss1987,Harrison1993,Haanappel1999,Pricopi2001,Bao2000,Fobes2017,Cornelius2001,Correa2005,Raymond2007,Fobes2018,Kanda2020,Hall2001a}

Recent results obtained at high magnetic fields suggested a unique behavior in CeRhIn$_5$. A field-induced QCP was reported to occur at the critical field $B_c \simeq 50$~T applied along both the $c$ and $a$ axes~\cite{Jiao2015,Jiao2019}. Furthermore, an electronic-nematic phase transition was observed at $B^{*}$, and attributed to an in-plane symmetry breaking~\cite{Ronning2017,Rosa2019}. While $B^*$ is only weakly temperature and angle dependent, its values obtained from different measurements vary from 27 to 31~T~\cite{Jiao2015,Moll2015,Ronning2017,Rosa2019,Lesseux2020,Kurihara2020}. Finally, Jiao \textit{et al}.~\cite{Jiao2015,Jiao2017} reported the emergence of additional dHvA frequencies at $B^{*} \approx$ 30~T, which was interpreted as a field-induced FS reconstruction associated with the $f$-electron delocalization. This result is surprising given that magnetic fields are generally expected to localize $f$ electrons. This motivated us to thoroughly reexamine the FSs of CeRhIn$_5$  at high magnetic fields.

In this Letter, we present the results of high-resolution angle-dependent dHvA measurements on CeRhIn$_5$ performed in static fields up to 36~T and pulsed fields up to 70~T. We find that several dHvA frequencies emerge at high fields but are not related to an FS reconstruction. We demonstrate that the $f$ electrons in CeRhIn$_{5}$ remain localized up to 70~T, well above the critical field to suppress the AFM order.

\begin{figure}[t]
\includegraphics[width=\columnwidth]{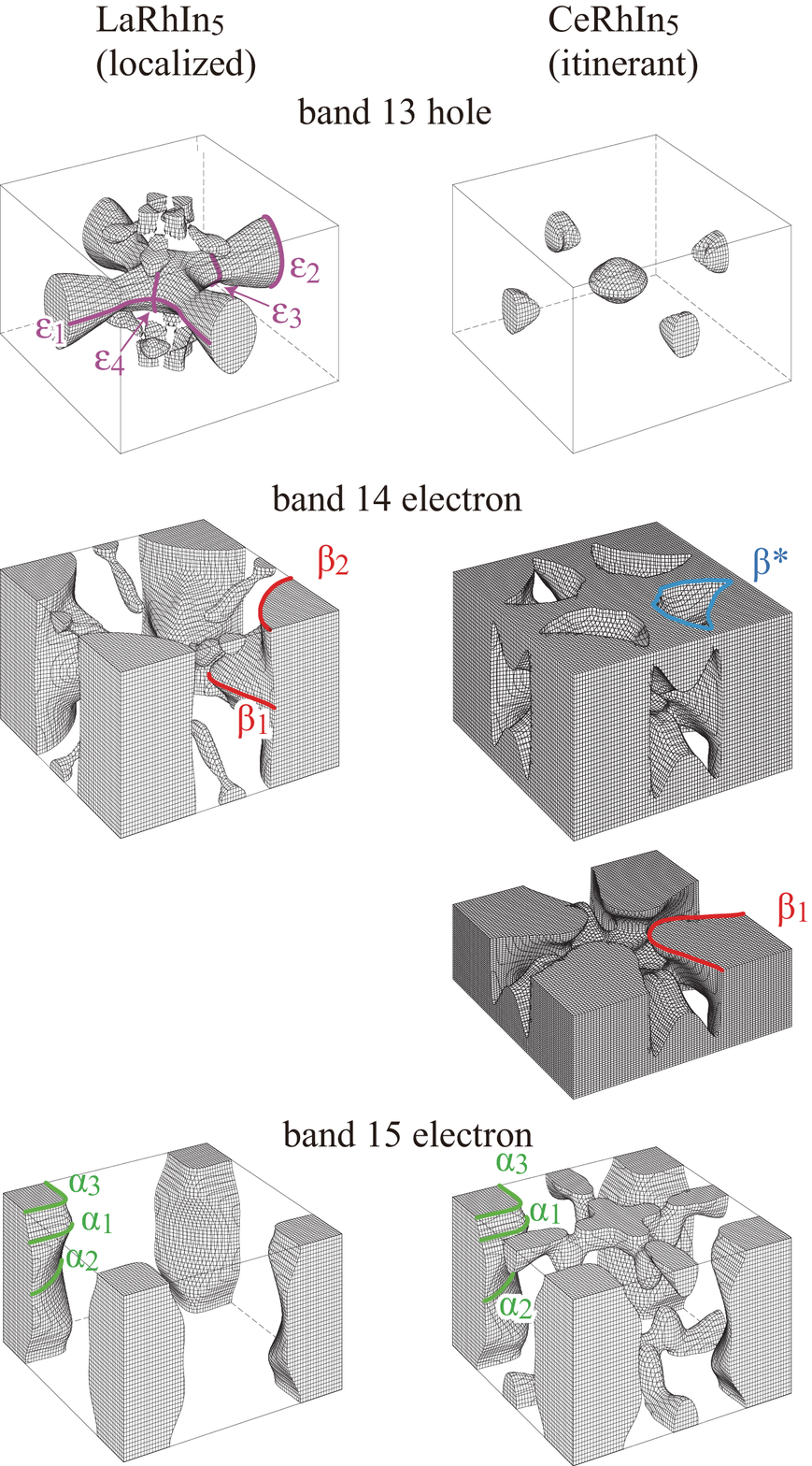}
\caption{\label{fig:FS}Calculated FSs with localized (left) and itinerant (right) $f$ electrons. Solid lines indicate some extremal cross sections as discussed in the text.}
\end{figure}

\begin{figure*}[ht]
\includegraphics[width=\textwidth]{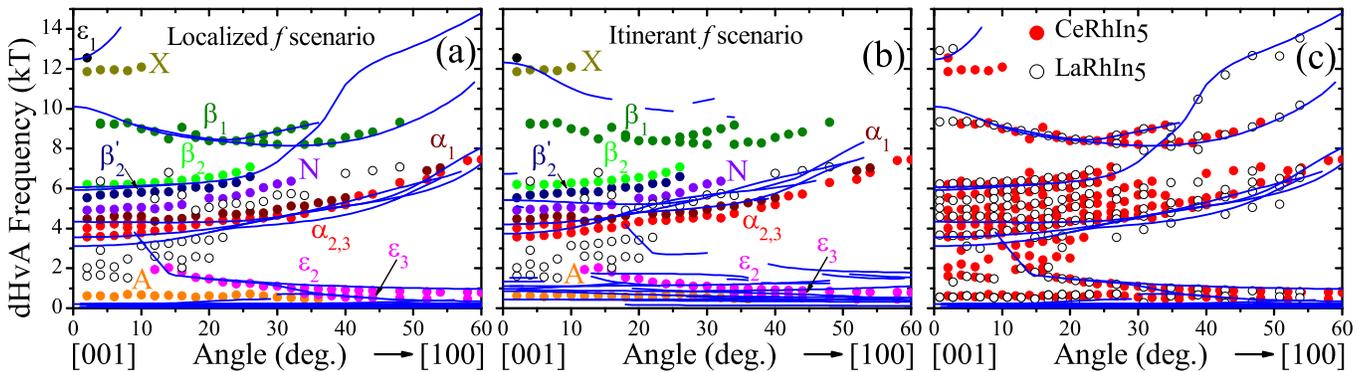}
\caption{\label{fig:ADL}Angular dependence of the experimentally observed dHvA frequencies in CeRhIn$_{5}$ together with results of band-structure calculations (solid lines) within localized (a) and itinerant (b) scenarios, and with the experimentally observed dHvA frequencies in LaRhIn$_{5}$ (open circles) (c). In the latter case, the band-structure calculations for localized $f$ electrons are also shown.}
\end{figure*}

Figure~\ref{fig:ADofQOFFT} shows the oscillatory torque after subtracting a nonoscillating background and the corresponding fast-Fourier transforms (FFTs) in CeRhIn$_{5}$ for several magnetic-field orientations $\theta$, where $\theta$ is the angle from the $c$ toward the $a$ axis. The FFTs are performed over a field interval mostly above $B^{*}$ with a high enough resolution to distinguish various dHvA frequencies. We observed all dHvA frequencies detected in previous low-~\cite{Hall2001,Shishido2002,Harrison2004} and high-field~\cite{Jiao2015,Jiao2017,Rosa2019} measurements. In addition, we observed a new previously undetected, dHvA frequency $\beta_1$, whose origin will be discussed later in more detail. Almost all of these frequencies are also observed in LaRhIn$_5$~\cite{Suppl}.

In order to address the question of whether the $f$ electrons in CeRhIn$_5$ are itinerant or localized at high magnetic fields, we performed band-structure calculations~\cite{Suppl} for both CeRhIn$_5$ with itinerant $f$ electrons and LaRhIn$_5$. Similar calculations were previously performed for CeCoIn$_5$~\cite{Settai2001,Shishido2002} and CeIrIn$_5$~\cite{Haga2001} with itinerant $f$ electrons and provided an excellent agreement with experimental results. The calculated FSs, shown in Fig.~\ref{fig:FS}, closely resemble those reported previously~\cite{Hall2001,Shishido2002,Elgazzar2004,Jiao2015}. Although there are similarities between the FSs with localized and itinerant $f$ electrons, several significant morphological differences between the two scenarios are apparent. The morphologies of the quasi-two-dimensional (2D) FS sheets originating from band 15 and giving rise to the $\alpha$ orbits are almost identical. Only the orbit size is different. There is, however, an additional three-dimensional (3D) FS sheet in the itinerant scenario. The FS sheets originating from band 13 are considerably different in the localized and itinerant cases. The complicated crosslike sheet giving rise to the $\varepsilon$ orbits in the localized case is replaced by two small ellipsoidal pockets in the itinerant scenario. The most essential difference, however, is the morphology of the FS sheets originating from the electron band 14. In the localized case, a quasi-2D sheet gives rise to the orbits $\beta_{1}$ (belly) and $\beta_{2}$ (neck). In the itinerant scenario, on the contrary, this sheet is more 3D with a larger $\beta_1$ orbit that is present over a more limited angular range and an entirely missing neck orbit $\beta_{2}$. The latter is replaced by a much smaller hole orbit $\beta^*$, which exists at very small angles only. This difference alone is sufficient to decide which of the calculated FSs yields a better agreement with the experimental results.

With this in mind, we will now compare the experimentally obtained angular dependence of the dHvA frequencies in CeRhIn$_{5}$ to the band-structure calculations for both localized [Fig.~\ref{fig:ADL}(a)] and itinerant [Fig.~\ref{fig:ADL}(b)] $f$ electrons. For this comparison, the dHvA frequencies were extracted from FFTs, such as shown in Fig.~\ref{fig:ADofQOFFT}, performed over the field range 29--36~T~\footnote{This range, mostly above $B^*$, was chosen to provide a high enough resolution to distinguish close dHvA frequencies. A small variation of this range does not alter the results.} ($B_{avg} =$~32.12~T~\footnote{Throughout this Letter, the average field, $B_{avg}$, of a field interval from $B_{min}$ to $B_{max}$ is defined as the reciprocal average, i.e., $1/B_{avg} = 1/2 (1/B_{min} + 1/B_{max})$.}).

There is excellent agreement between the experimentally observed branches $\alpha_{2,3}$, $\alpha_1$, $\beta_2$, and $\beta_1$ and their counterparts calculated for localized $f$ electrons [Fig.~\ref{fig:ADL}(a)]. These branches originate from the quasi-2D FS sheets from bands 14 and 15. The experimentally observed branches $\varepsilon_2$ and $\varepsilon_3$ also agree very well with the $f$-localized calculations. These branches originate from the crosslike FS sheet from band 13. The branch $X$ is observed only at small angles close to the $c$ axis, and it does not seem to correspond to any calculated extremal area. Neither do the branches $\beta_{2}'$ and $N$ correspond to any calculated branch. These frequencies will be discussed later in more detail.

On the other hand, there is a clear disagreement of the experimental data with band-structure calculations assuming itinerant $f$ electrons [Fig.~\ref{fig:ADL}(b)]. The most significant signature of this disagreement is the presence of the $\beta_{2}$ branch, which is supposed to be completely absent in the itinerant scenario. This branch was also observed in previous high-field dHvA~\cite{Jiao2015,Jiao2017} and magnetostriction~\cite{Rosa2019} measurements. In measurements under high pressure, this frequency disappears above $P_c$, where the $f$ electrons become itinerant~ \cite{Shishido2005}. Furthermore, the experimentally observed $\alpha_{2,3}$, $\alpha_1$, and $\beta_1$ branches lie far below those calculated for the itinerant case. In previous high-field dHvA measurements performed with a magnetic field only along the $c$ axis, the frequencies $N$ and $\beta_{2}'$ were interpreted as $\alpha_2$ and $\alpha_1$ of the $f$-itinerant model~\cite{Jiao2015}. However, their angular dependence is clearly different from that suggested by the calculations. Similarly, the experimentally observed branch $X$ was interpreted as the $\beta_1$ branch of the itinerant scenario~\cite{Jiao2015}. This frequency indeed agrees with calculated $\beta_1$ values at small angles close to the $c$ axis. The positive curvature of this branch, however, is inconsistent with the negative curvature of the calculated $\beta_1$ branch.

\begin{figure}[ht]
\includegraphics[width=\columnwidth]{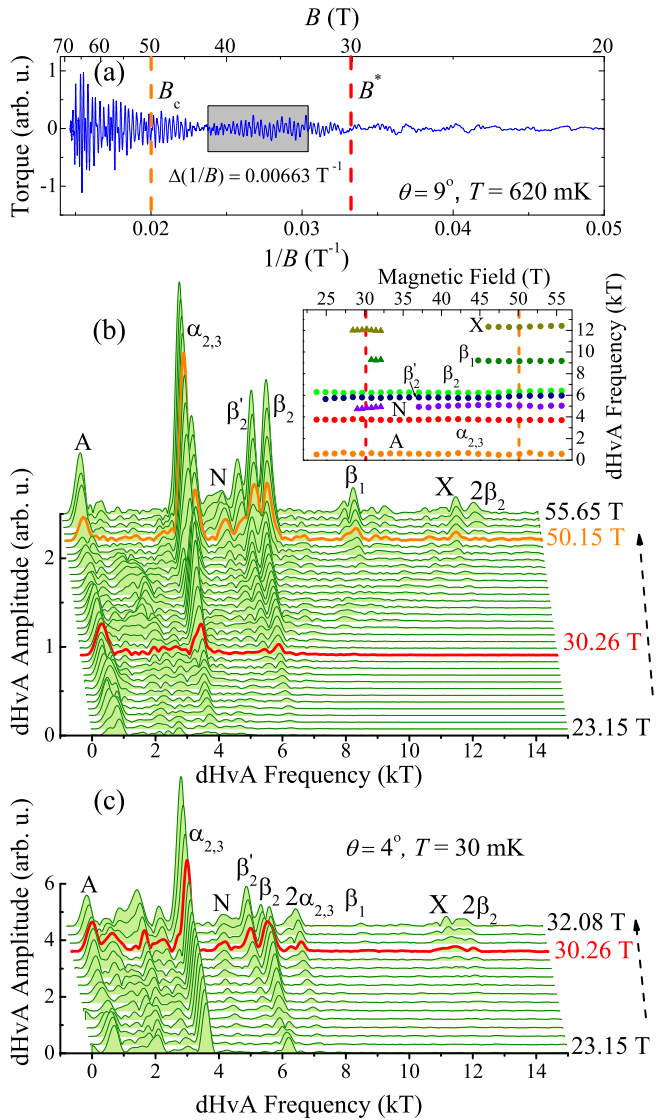}
\caption{\label{fig:AFM} dHvA oscillations in CeRhIn$_{5}$ (a) in pulsed magnetic fields applied at 9$^\circ$ from the $c$ axis. (b) FFT spectra of the oscillations from (a) obtained over the same $1/B$ range, shown by the rectangle in (a). For the bottom curve, the range is from $B_{min}$ = 21.5~T to $B_{max}$ = 25.07~T ($B_{avg}$ = 23.15~T). For each successive curve, $B_{min}$ is increased by 0.5~T up to 29~T and by 1~T from there on. The inset shows the evolution of the dHvA frequencies with $B$ obtained from pulsed (circles) and static (triangles) field measurements. (c) FFT spectra of the static-field dHvA oscillations, shown in Fig.~\ref{fig:ADofQOFFT}, with $B$ at 4$^\circ$ from the $c$ axis. The $1/B$ range and the range for the bottom curve are the same as in (b). For each successive curve, $B_{min}$ is increased by 0.5 T. The curves in (b) and (c) are shifted for clarity.}
\end{figure}

Finally, in Fig.~\ref{fig:ADL}(c), we compare the angular dependence of the experimentally observed dHvA frequencies in CeRhIn$_5$ and LaRhIn$_5$. The dHvA frequencies of the latter compound were also obtained from high-field ranges up to the available $B_{max}$~\cite{Suppl}. The excellent agreement of the experimentally measured dHvA frequencies observed in CeRhIn$_5$ and LaRhIn$_5$ over the whole angular range is an especially strong confirmation that the $f$ electrons, which are localized at low fields~\cite{Shishido2002}, remain so even in fields higher than $B^*$. Furthermore, the dHvA frequencies observed in CeRhIn$_5$ at high fields do not match those observed above the critical pressure~\cite{Suppl}.

Figures~\ref{fig:ADofQOFFT} to~\ref{fig:ADL} convincingly demonstrate that the Ce $f$ electrons are localized up to 36~T. That this situation  persists up to still higher fields is evidenced by our pulsed-field measurements, shown in Fig.~\ref{fig:AFM}(a). The corresponding FFT spectra obtained over a moving $1/B$ window are shown in Fig.~\ref{fig:AFM}(b). Due to the much higher temperature, $T =$~620~mK, of the pulsed-field measurements, some of the additional dHvA frequencies, such as $N$, $\beta_1$, and $X$, emerge at higher fields as compared to our lower temperature static-field results, an example of which is shown in Fig.~\ref{fig:AFM}(c). Contrary to the previous report~\cite{Jiao2015}, some of these frequencies emerge well below $B^*$, as can be seen in Fig.~\ref{fig:AFM}(b) and (c). We have not observed any frequency shift or the emergence of new frequencies at either $B^*$ or $B_c$ (inset of Fig.~\ref{fig:AFM}). Importantly, the $\beta_2$ frequency is still present above $B_c \simeq$~50~T. Furthermore, the effective masses remain finite in the immediate vicinity of $B_c$~\cite{Suppl}. This suggests that the $f$ electrons remain localized when the AFM order is suppressed by a magnetic field. The same conclusion was drawn for CeIn$_3$, for which the AFM order is also suppressed at a very high field of about 60~T~\cite{Harrison2007}.

We will now discuss the origin of the additional dHvA frequencies, such as $\beta_{2}'$, $N$, $\beta_{1}$, and $X$, which emerge only at high fields. The frequency $\beta_{1}$ originates from the belly orbit of the electron band 14 (Fig.~\ref{fig:FS}). Both the frequency itself and its angular dependence are in excellent agreement with the $f$-localized calculations [Fig.~\ref{fig:ADL}(a)]. This branch was not observed in any of the previous measurements. This is not surprising given that this orbit is strongly affected by the modification of the Brillouin zone in the AFM state~\cite{Suppl}. In addition, the corresponding effective mass is strongly enhanced, $m^* \simeq 12~m_0$~\cite{Suppl}. The frequencies $\beta_{2}'$ and $N$ do not seem to correspond to any calculated branch within either the localized or itinerant model. It is, however, apparent that $\beta_{2}'$ follows the angular variation of $\beta_{2}$ with a small and almost constant offset (Fig.~\ref{fig:ADL}). Remarkably, this offset $\sim0.6$~kT, corresponds to the frequency $A$, which is almost angle independent. This suggests that $\beta_{2}'$ originates from the magnetic breakdown between $\beta_{2}$ and $A$, i.e., $\beta_{2}' = \beta_{2} - A$, assuming that $A$ corresponds to a hole pocket. The exact origin of the frequencies $N$ and $X$ is unclear at present. Notably, the $\beta_{2}'$ and $N$ branches also emerge in LaRhIn$_5$ only at high fields [(see Fig.~\ref{fig:ADL}(c)]~\cite{Suppl}, suggesting that they originate from magnetic-breakdown orbits. The frequency $X$ is the only one that is not observed in LaRhIn$_5$. It is, therefore, less likely that this frequency originates from a magnetic breakdown. Its emergence at high fields, although progressive~\cite{Suppl}, could be due to a minor FS reconstruction, such as a Lifshitz transition of a spin-split band, similar to what was observed in CeIrIn$_5$~\cite{Aoki2016} and YbRh$_2$Si$_2$~\cite{Rourke2008}. We emphasize that even if this is the case, the $f$ electrons remain localized above 30~T, as is evidenced by the presence of the $\beta_1$ and $\beta_2$ frequencies, and the very close match between the LaRhIn$_5$ and the CeRhIn$_5$ data.

In summary, we performed high-field dHvA measurements on CeRhIn$_5$ and LaRhIn$_5$. In CeRhIn$_5$, several additional dHvA frequencies emerge above certain threshold fields. In particular, we observed the previously undetected, thermodynamically important $\beta_1$ branch predicted by the $f$-localized band-structure calculations. Almost all of the dHvA frequencies observed in CeRhIn$_5$ are also present in LaRhIn$_5$. In addition, their angular dependence is identical in the two compounds. The presence and angle dependence of the observed dHvA frequencies are well accounted for by band-structure calculations with localized $f$ electrons,  indicating that the $f$ electrons of CeRhIn$_{5}$ remain localized not only above $B^* \approx$ 30~T, but also above $B_c \simeq$ 50~T. We emphasize that delocalization of the Ce $f$ electron at high magnetic field would change the whole FS from the localized to the itinerant FS that we show in Fig.~\ref{fig:FS}.  Continued observation of the $\beta_2$ branch at the highest magnetic fields is clear evidence of that the Ce $f$ electron remains localized over the full magnetic field range that we have explored.

It was previously reported that the $f$ electrons also remain localized in CeIn$_3$ above its critical field $B_c \simeq$~60~T~\cite{Harrison2007}. Whereas CeIn$_3$ is an isotropic HF compound with an almost spheroidal FS, CeRhIn$_5$ is a prototypical example of a strongly anisotropic material with quasi-2D FSs. The continued localization of the $f$ electrons well above $B_c$ in both compounds is not consistent with either of the two existing theoretical models of AFM QCPs. This implies that magnetic field, which itself tends to localize $f$ electrons, should be treated differently from such control parameters as pressure or chemical doping.

\begin{acknowledgments}
We thank H. Shishido for sharing with us the high-pressure dHvA data and R. Settai for sharing with us the results obtained in CeCoIn$_5$. We acknowledge the support of the LNCMI-CNRS, the HLD-HZDR, and the HFML-RU, members of the European Magnetic Field Laboratory (EMFL), the ANR-DFG grant ``Fermi-NESt,'' the DFG through the excellence cluster $ct.qmat$ (EXC 2147, Project ID 39085490), and JSPS KAKENHI Grants Nos. JP15H05882, JP15H05884, JP15H05886, and JP15K21732 (J-Physics).
\end{acknowledgments}

\bibliography{CeRhIn5_dHvA}

\begin{thebibliography}{57}
\bibitem [{Sup()}]{Suppl}%
  \BibitemOpen
  \href@noop {} {\bibinfo {title} {See {S}upplemental {M}aterial, which
  includes {R}efs. {[24--42]}, at [{URL}] for the experimental details, {dHvA}
  oscillations in {LaRhIn$_5$}, the details of the band-structure calculations,
  effective masses, angle-dependence of the metamagnetic transition,
  field-dependence of the oscillatory amplitudes of the {$X$} and $\beta_1$
  branches, and comparison of the {dHvA} frequencies in {CeRhIn$_5$} at high
  magnetic fields and under high pressure.}}\BibitemShut {Stop}%
\end{thebibliography}

\clearpage

\section*{Supplemental Material for ``Robust Fermi-Surface Morphology of CeRhIn$_5$ across the Putative Field-Induced Quantum Critical Point''}

\subsection{Methods}

High-quality (residual resistivity ratio $\rho_{RT}/\rho_0 \gtrsim$ 300) single crystals of CeRhIn$_{5}$ were grown by the In self-flux method, details of which are given elsewhere~\cite{Shishido2002}. The dHvA experiments were performed using magnetic-torque methods in both static (up to 36~T) and pulsed (up to 70~T) magnetic fields. The former were performed in a dilution refrigerator ($T_{base} \sim$ 30~mK) equipped with a low-temperature rotator using a metallic capacitive cantilever. The latter were done in a $^{3}$He cryostat ($T_{base} \sim$ 620~mK) using a piezoresistive microcantilever.

\begin{figure}[htb]
\includegraphics[width=\columnwidth]{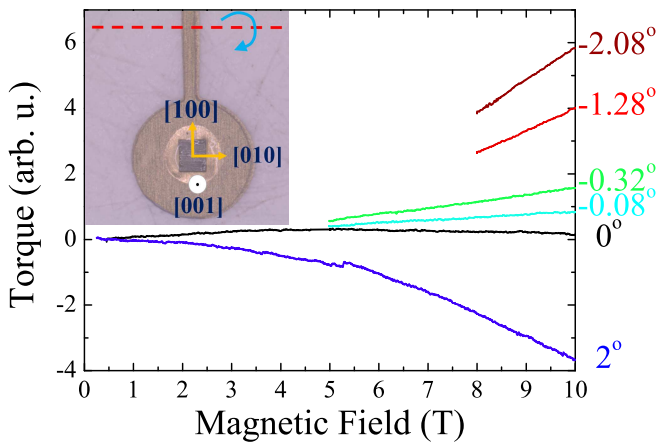}
\caption{\label{fig:Alignment}Magnetic torque as a function of magnetic field applied at small angles close to the $c$ axis. The inset shows a photograph of the CeRhIn$_5$ sample mounted on a metallic cantilever.}
\end{figure}

When using a rotator, the sample was first carefully aligned on the cantilever with respect to the rotation axis, as shown in the inset of Fig.~\ref{fig:Alignment}. The whole setup was then mounted in the rotator of the dilution refrigerator probe at a small negative angle. The sample was then rotated in field in small steps until the background torque vanished, as shown in Fig.~\ref{fig:Alignment}. Since magnetic torque vanishes when a magnetic field is applied along a symmetry axis, this orientation was taken as the reference, i.e., field parallel to the $c$ axis. The estimated error in the angle determination is less than 0.1$^\circ$. Once the reference orientation was found, the sample was always rotated in the same direction to avoid hysteresis.

\subsection{dHvA oscillations in LaRhIn$_5$}

\begin{figure}[htb]
\includegraphics[width=\columnwidth]{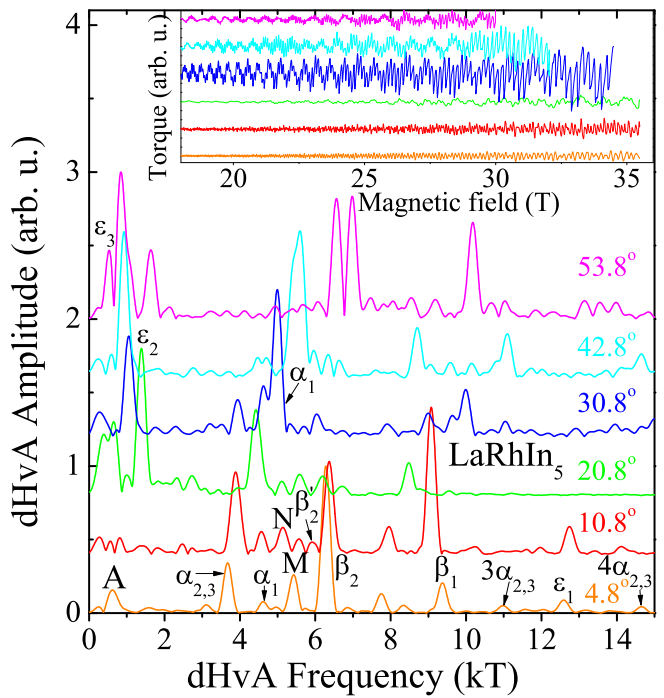}
\caption{\label{fig:QOLa}Fourier spectra of the dHvA oscillations (shown in the insets) of LaRhIn$_5$ for magnetic field applied at various angles, $\theta$, from the $\textit{c}$ towards the $\textit{a}$ axis at $T =$~50~mK. The curves are shifted vertically for clarity. The FFTs in are performed over the same $1/B$ interval, $\Delta (1/B) \approx$ 0.028~T$^{-1}$, with $B_{min}$ adjusted with respect to the available $B_{max}$  (see the inset) due to experimental constraints. All FFTs are normalized to the strongest dHvA spectral peak.}
\end{figure}

\begin{figure*}[ht]
\includegraphics[width=\textwidth]{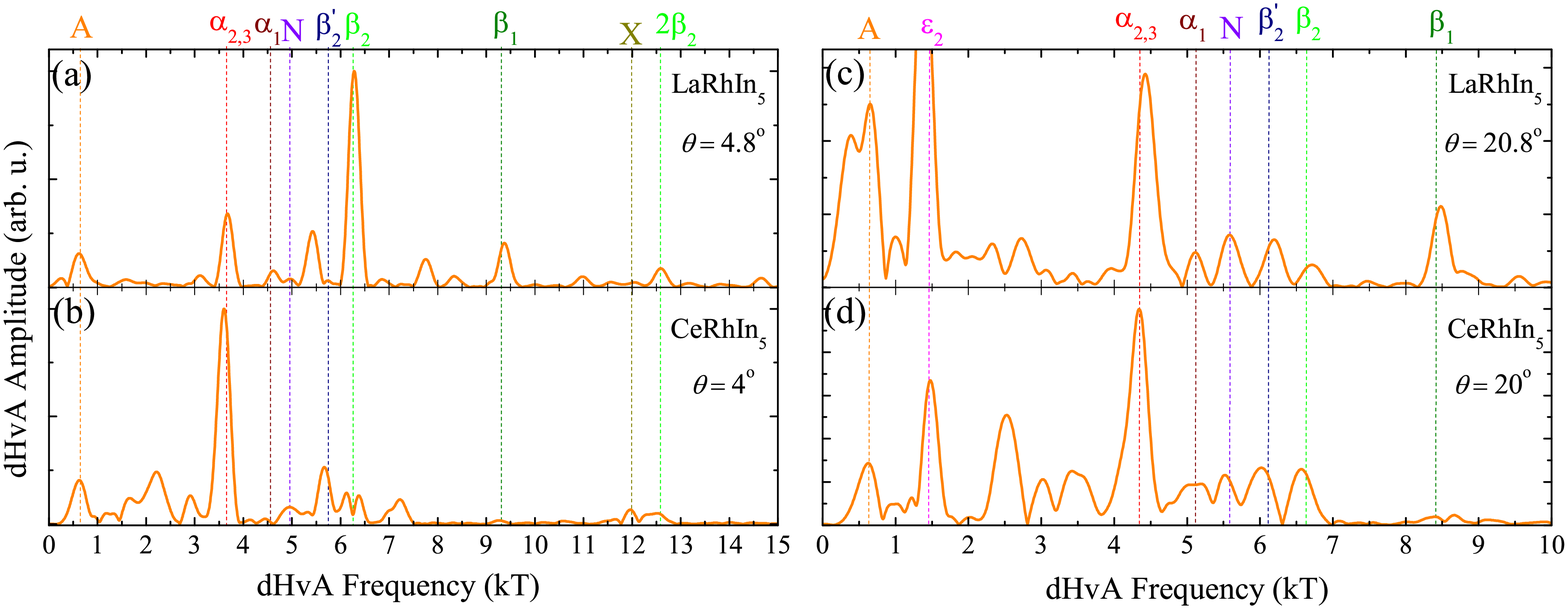}
\caption{\label{fig:FFTcomparison}Comparison of the high field FFT spectra of the dHvA oscillations in LaRhIn$_5$ [(a) and (c)] and CeRhIn$_5$ [(b) and (d)] with magnetic applied at two close angles of 4.8$^\circ$ (a) and 4$^\circ$ (b), and 20.8$^\circ$ (c) and 20$^\circ$ (d).}
\end{figure*}

\begin{figure}[htb]
\includegraphics[width=\columnwidth]{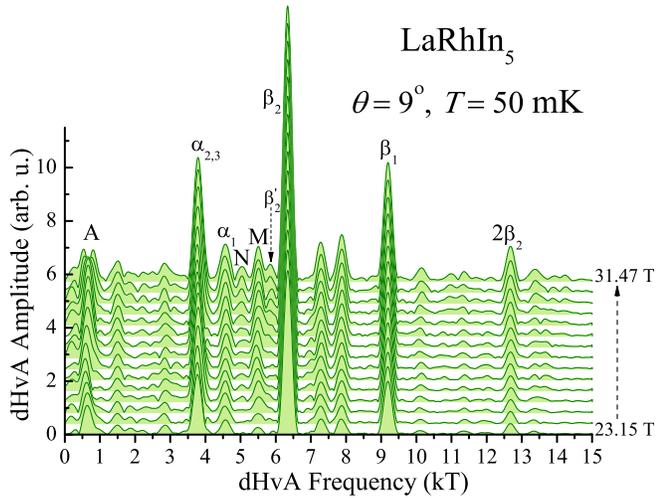}
\caption{\label{fig:FvsBLa}FFT spectra of the dHvA oscillations in LaRhIn$_5$ with magnetic field applied at 9$^\circ$ from the $c$ axis at $T =$ 50 mK. All the FFTs are obtained over the same $1/B$ range. For the bottom curve, the range is from $B_{min}$ = 21.5~T to $B_{max}$ = 25.07~T ($B_{avg}$ = 23.15~T). For each successive curve, $B_{min}$ is increased by 0.5 T. The curves are shifted vertically for clarity.}
\end{figure}

Figure~\ref{fig:QOLa} shows the oscillatory torque after subtracting a non-oscillating background and the corresponding FFTs in LaRhIn$_{5}$ for several magnetic-field orientations $\theta$, where $\theta$ is the angle from the $c$ towards the $a$ axis. We observed all the fundamental dHvA frequencies, namely $\alpha_i$ and $\beta_i$, predicted by the band-structure calculations. These frequencies were also observed in previous measurements in fields up to 13~T~\cite{Shishido2002}. All these frequencies were also observed in CeRhIn$_5$, as shown in Fig.~\ref{fig:FFTcomparison}. In addition to these fundamental frequencies, several additional frequencies, such as $N$, $M$, and $\beta_{2}'$, gradually emerge at high fields, as shown in Fig.~\ref{fig:FvsBLa}. These frequencies were not observed in previous lower field measurements~\cite{Shishido2002}. This suggests that they originate from magnetic breakdown orbits. Remarkably, the frequencies $N$, and $\beta_{2}'$, were also observed to emerge at high fields in CeRhIn$_5$, as shown in Fig.~\ref{fig:FFTcomparison} and discussed in the main text.

\subsection{Band-structure calculations}

Band-structure calculations were carried out using a full potential linearized augmented plane wave (FLAPW) method with the local density approximation (LDA) for the exchange correlation potential. For the LDA, the formula proposed by Gunnarsson and Lundqvist~\cite{Gunnarsson1976} was used. For the band-structure calculations, the program codes \textsc{tspace} and \textsc{kansai} were used.

The space group of CeRhIn$_5$ is $P4/mmm$ (\# 123, $D_{4h}^{1}$). The lattice parameters used for the calculation are $a=4.6521$~\AA~and $c=7.5404$~\AA~\cite{Haga}. These parameters are similar to those previously reported~\cite{Hegger2000,Thompson2001,Moshopoulou2001,Moshopoulou2002}. In $P4/mmm$, the $1a$ (0.0, 0.0, 0.0) and $1b$ site (0.0, 0.0, 0.5) are occupied by Ce and Rh ions, respectively. Indium ions occupy the $1c$ (0.5, 0.5, 0) and $4i$ site (0.0, 0.5, 0.3068). In the calculation for LaRhIn$_5$, the Ce ion was just replaced by the La ion, because this is a non-$4f$ reference material.

In the FLAPW method, the scalar relativistic effect~\cite{Koelling1977} is considered for all electrons and the spin-orbit coupling is included self-consistently for all valence electrons as a second variational procedure. The muffin-tin (MT) sphere radii are set as 0.3543$a$ for Ce and La, and 0.32806$a$ for Rh and In. Here, $a$ is the lattice constant for the $a$ axis. Core electrons (Xe core minus 5$s^2$5$p^6$ for Ce and La, Kr core minus 4$p^6$ for Rh, Kr core for In) are calculated inside the MT sphere in each self-consistent step. 5$s^2$5$p^6$ electrons on Ce and La, 4$p^6$ on Rh, and 4$d^{10}$ on In ions are calculated as valence electrons by using the second energy window. The LAPW basis functions are truncated at $|{\bf k+G}_i| \leq 4.85$ $(2\pi/a)$, corresponding to 771 LAPW functions at the $\Gamma$ point. 225 sampling points uniformly distributed in the irreducible 1/16th of the Brillouin zone (BZ) (2048 points in the full BZ) were used for potential convergence and the final band structure.

\begin{figure}[htb]
  \includegraphics[width=\columnwidth]{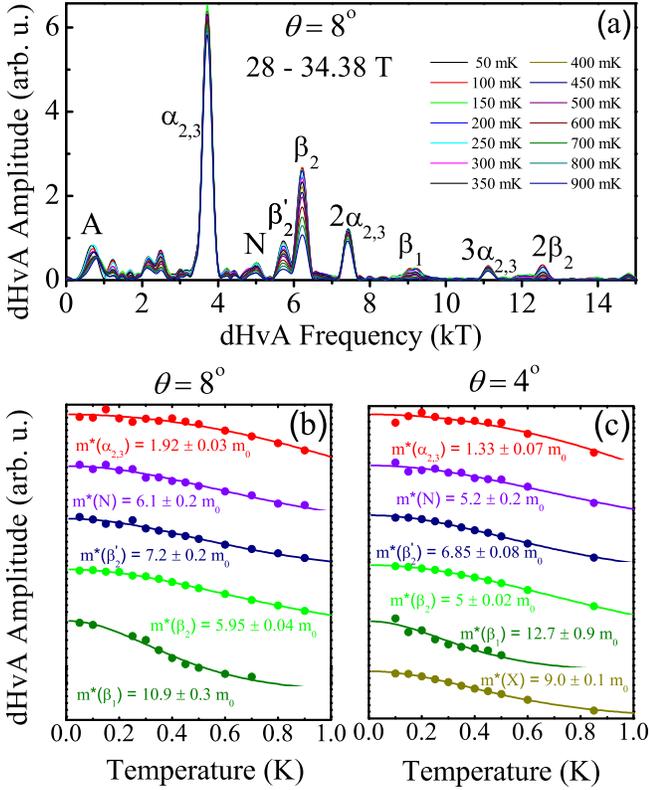}
  \caption{(a) Fourier spectra of the dHvA oscillations in CeRhIn$_5$ for a field rotated by $\theta = 8^\circ$ from $c$ to the $a$ axis over the field interval from 28 to 34.5 T at different temperatures. (b) and (c) Temperature dependence of the dHvA amplitudes at $\theta = 8^\circ$ and 4$^\circ$, respectively. The lines are fits using the standard Lifshitz-Kosevich formula~\cite{Shoenberg2009}.}\label{fig:Massplot}
\end{figure}

\subsection{Effective masses in CeRhIn$_{5}$}

\begin{figure}[htb]
  \includegraphics[width=\columnwidth]{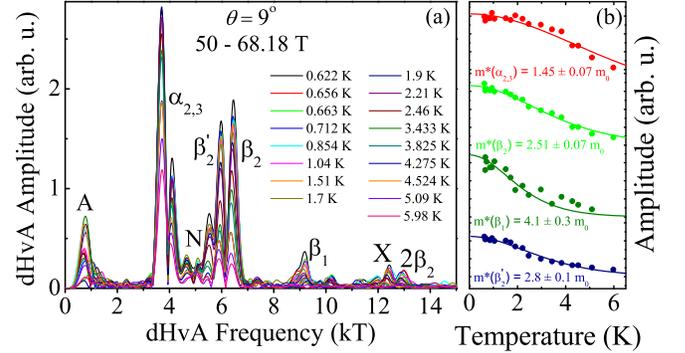}
  \caption{(a) FFT spectra of the dHvA oscillations in CeRhIn$_5$ for a field rotated by $\theta = 9^\circ$ from $c$ to the $a$ axis over the field interval from 50 to 68.18 T at different temperatures. (b) Temperature dependence of the dHvA amplitudes from (a). The lines are fits using the standard Lifshitz-Kosevich formula~\cite{Shoenberg2009}.}\label{fig:Massplotpulsed}
\end{figure}

The effective masses, $m^*$, corresponding to the various dHvA orbits of CeRhIn$_5$ were determined by fitting the temperature dependence of the oscillatory amplitude with the standard Lifshitz-Kosevich formula~\cite{Shoenberg2009} with $m^\star$ as fitting parameter. This was done deep inside the antiferromagnetic phase in the vicinity of $B^*$ (Fig.~\ref{fig:Massplot}) and in the field-induced polarized paramagnetic phase just above $B_c$ (Fig.~\ref{fig:Massplotpulsed}).

Inside the antiferromagnetic state, the effective masses were determined from the static-field data over the field range from 28 to 34.5~T ($B_{avg}$ = 30.86~T) for two orientations of the magnetic field, at 4$^\circ$ and 8$^\circ$ from $c$ towards the $a$ axis, as shown in Fig.~\ref{fig:Massplot}. The effective masses obtained for these two orientations are similar, indicating a rather weak angular dependence of the effective masses, at least close to the $c$ axis. The effective masses corresponding to the dHvA frequencies observed at moderate magnetic fields, such as $\alpha_{2,3}$ and $\beta_2$, are in good agreement with previous reports~\cite{Hall2001,Shishido2002}. The effective masses corresponding to the additional frequencies observed only at high fields are much heavier, from $\sim$6$m_0$ for the branch $N$, to $\sim$12$m_0$ for the branch $\beta_1$ (where $m_0$ is the bare electron mass), indicating many-body mass enhancements. For comparison, the calculated band masses, $m_b$, for the $\beta_1$ branch are 0.85$m_0$ and 0.81$m_0$ at 4$^\circ$ and 8$^\circ$, respectively.

In the field-induced polarized paramagnetic phase, the effective masses were extracted from the pulsed-field data over the field range from 50 to 68.18~T ($B_{avg}$ = 57.69~T) with field applied at 9$^\circ$ from $c$ towards the $a$ axis, as shown in Fig.~\ref{fig:Massplotpulsed}. Not only do the effective masses show no sign of divergence just above $B_c$, the effective masses of the $\beta_2$ and $\beta_1$ branches are strongly reduced with respect to the values at 30.86~T. Unfortunately, the amplitudes of these oscillations are not strong enough for a reliable determination of the effective masses in the intermediate field range. The apparent reduction of the effective masses observed here is likely to be due to the field-induced polarization of the quasiparticle bands, which leads to the suppression of electronic correlations. A similarly strong suppression of the effective masses by magnetic fields was previously observed in CeB$_6$~\cite{Joss1987,Harrison1993}, CeAl$_2$~\cite{Haanappel1999,Pricopi2001}, CeRu$_2$Si$_2$~\cite{Pricopi2001}, and CeCoIn$_5$~\cite{Settai2001}.

\subsection{Field-dependence of the first-order phase transition from incommensurate to commensurate magnetic structure}

\begin{figure}[htb]
\includegraphics[width=\columnwidth]{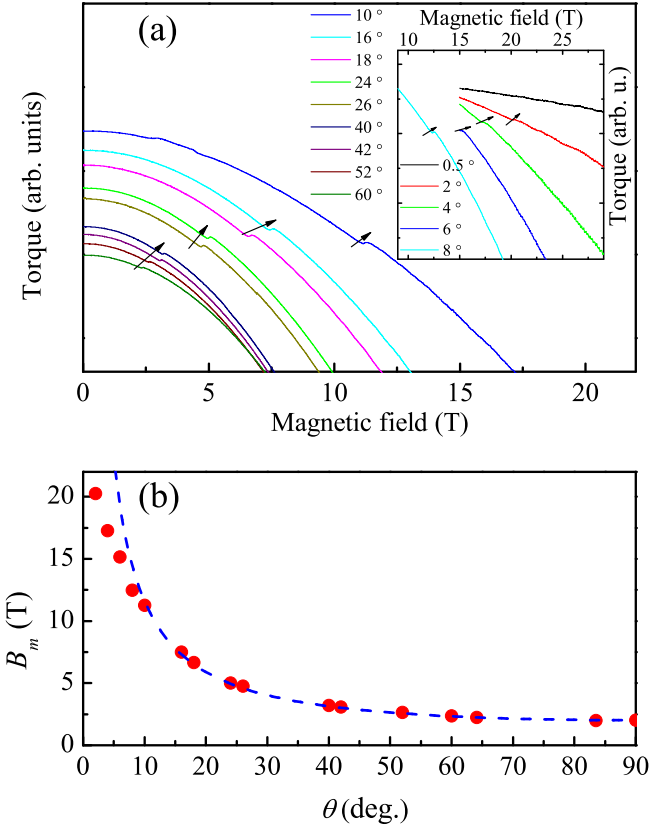}
\caption{First-order metamagnetic transition in CeRhIn$_{5}$. Black arrows in (a) indicate the metamagnetic transition in the raw torque signal at various field orientations, $\theta$. (b) Angular dependence of the metamagnetic transition field $B_{m}$. The dashed line represents $B_m^a / \cos(\alpha) = B_m^a / \sin(\theta)$, where $B_m^a = 2$~T is the transition field along the $a$ axis, and $\alpha$ is the angle from $a$ to the $c$ axis.} \label{fig:metamagnetic}
\end{figure}

At ambient pressure and zero magnetic field, the magnetically ordered ground state of CeRhIn$_5$ is an incommensurate helicoidal phase characterized by the propagation vector $\mathbf{k} = (1/2, 1/2, 0.297)$~\cite{Bao2000}. The magnetic moments on the cerium ions are aligned antiferromagnetically in the tetragonal basal plane and spiral transversely along the $c$ axis~\cite{Bao2000,Fobes2017}. When the magnetic field is applied in the basal plane, a first-order phase transition occurs at $B_m \simeq 2$~T at low temperatures~\cite{Shishido2002,Cornelius2001,Correa2005}. The transition corresponds to a change of the magnetic structure to a commensurate sine-wave structure with the propagation vector $\mathbf{k} = (1/2, 1/2, 1/4)$~\cite{Raymond2007,Fobes2018,Kanda2020}. The magnetic moments are still in the basal plane, but now aligned perpendicular to the applied field direction~\cite{Raymond2007,Fobes2018}. When the magnetic field is tilted away from the basal plane, the transition field $B_{m}$ increases, and initially follows a $1 / \cos(\alpha)$ dependence, where $\alpha$ is the angle from the basal plane towards the $c$ axis~\cite{Shishido2002}.

In magnetic-torque measurements, the transition manifests itself by a clear anomaly, as shown in Fig.~\ref{fig:metamagnetic}(a). $B_{m}$ deviates from the $1 / \cos(\alpha)$ dependence at higher angles [see Fig.~\ref{fig:metamagnetic}(b)]. It occurs slightly above 20~T when the field is applied at $\theta = 2^\circ$, the lowest angle at which we observed the corresponding anomaly.

All the static-field measurements reported here were performed at fields higher than $B_m$, i.e., in the commensurate phase with $\mathbf{k} = (1/2, 1/2, 1/4)$. This implies that the magnetic Brillouin zone is smaller than its paramagnetic counterpart. As a consequence, the topology of the Fermi surface in the antiferromagnetic state is modified with respect to the paramagnetic phase, for which the band-structure calculations are performed. This might be approximated by a band-folding procedure where the paramagnetic Fermi surface is folded into the smaller Brillouin zone based on a large magnetic unit cell. In the particular case of CeRhIn$_5$, the reduction of the Brillouin zone hardly affects the Fermi surface sheet originating from band 15. On the other hand, the sheet originating from band 14, especially the largest $\beta_1$ orbit is strongly affected. This orbit can only be observed via magnetic breakdown.

\subsection{Field-dependence of the amplitudes of the dHvA frequencies $X$ and $\beta_1$}

\begin{figure}[htb]
\includegraphics[width=\columnwidth]{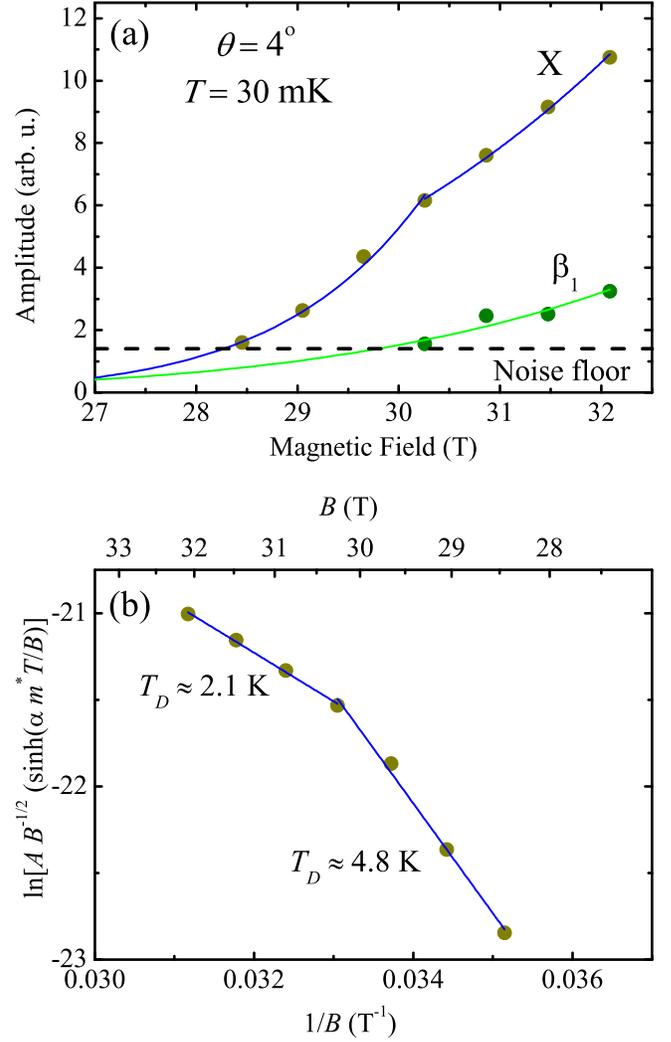}
\caption{\label{fig:Dingle} (a) dHvA amplitude of the $X$ and $\beta_1$ branches as a function of magnetic field applied at 4$^\circ$ from the $c$ axis at $T =$ 50~mK. The lines correspond to the standard Lifshitz-Kosevich formula~\cite{Shoenberg2009} with effective masses and Dingle temperatures determined from mass and Dingle plots, respectevely. (b) Dingle plot for the branch $X$. Note that the slope of the Dingle plot changes drastically at about 30~T.}
\end{figure}

Figure~\ref{fig:Dingle}(a) shows the field dependence of the oscillatory amplitude of the $X$ and $\beta_1$ branches. The lines represent the field-dependent part of the Lifhitz-Kosevich formula. The effective masses were obtained from the mass plots shown above, and the Dingle temperatures were determined from the Dingle plots, as shown in Fig.~\ref{fig:Dingle}(b). For both frequencies, the expected amplitude drops below the noise level just below the lowest field point. This implies that these oscillations gradually rise above the noise level with increasing magnetic field, rather than emerging all of a sudden at a given field value. Indeed, in the latter case, the oscillations are expected to emerge well above the noise level.

The kink in the field-dependence of the $X$ frequency amplitude is due to a drastic change of the Dingle temperature, as shown in Fig.~\ref{fig:Dingle}(b). Remarkably, this change occurs at about 30~T. While it is not clear whether this change is directly related to $B^*$, it might explain why the $X$ frequency was not observed below 30~T in the previous high field study~\cite{Jiao2015}.

\subsection{Comparison of the dHvA frequencies in CeRhIn$_5$ at high magnetic fields and under high pressure}

\begin{figure}[htb]
\includegraphics[width=\columnwidth]{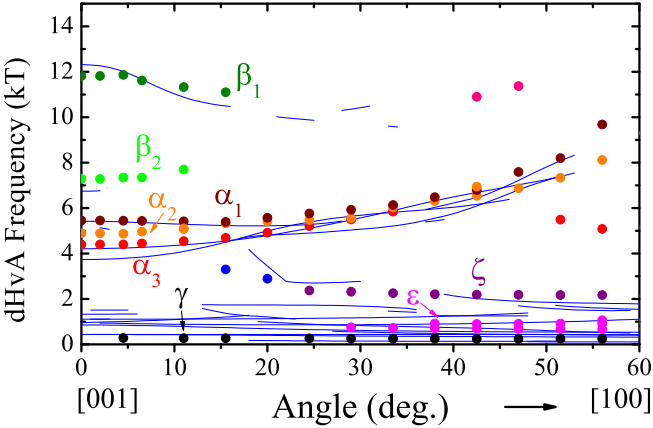}
\caption{\label{fig:CeCoIn5}Angular dependance of the experimentally observed dHvA frequencies in CeCoIn$_5$ (from Ref.~\cite{Settai2001}) together with the results of $f$-itinerant band-structure calculations for CeRhIn$_5$ (solid lines).}
\end{figure}

The heavy fermion compounds CeRhIn$_{5}$ and CeCoIn$_{5}$ are iso-structural with similar Fermi-surface topology. However, contrary to CeRhIn$_5$, CeCoIn$_{5}$ is non-magnetic with itinerant $f$ electrons~\cite{Settai2001,Hall2001a,Shishido2002}. That is why the two cases are often compared to establish whether the $f$ electrons of CeRhIn$_5$ are itinerant or localized under certain conditions, such as high pressure~\cite{Shishido2005} or high magnetic fields~\cite{Jiao2015,Jiao2017}. Strictly speaking, this approach is not correct as there are certain differences between the Fermi surfaces of CeCoIn$_5$ and CeRhIn$_{5}$ with itinerant $f$ electrons, as shown in Fig.~\ref{fig:CeCoIn5}. Not only the sizes of the $\alpha_2$ and $\alpha_3$ orbits are slightly different, the $\beta_2$ orbit, present in CeCoIn$_5$, is absent in CeRhIn$_{5}$ with itinerant $f$ electrons. The appropriate way to experimentally assess whether the $f$ electrons of CeRhIn$_5$ are itinerant or localized under certain conditions is to compare the data with those obtained in CeRhIn$_5$ above the critical pressure, $P_c =$ 2.3~GPa, where the $f$ electrons are known to be itinerant~\cite{Shishido2005}.

\begin{figure}[htb]
\includegraphics[width=\columnwidth]{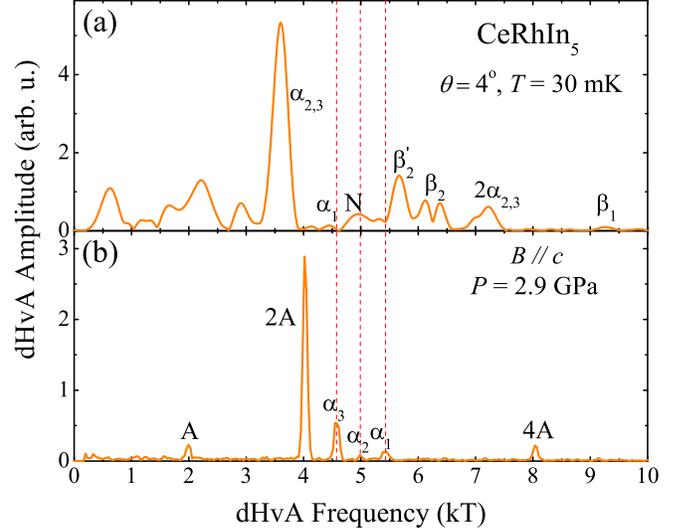}
\caption{\label{fig:HighPressure}Comparison of the dHvA frequencies in CeRhIn$_5$ at high magnetic fields (a) and under high pressure (b). The results under pressure are from Ref.~\cite{Shishido2005}.}
\end{figure}

In Fig.~\ref{fig:HighPressure}, we show a comparison of the dHvA frequencies in CeRhIn$_{5}$ over the field interval 29 - 36~T, mostly above B$^{*}$, with those at $P =$ 2.9~GPa, well above $P_c$~\cite{Shishido2005}. This comparison is not straightforward, as the measurements under pressure were performed with field applied along the $c$ axis, where dHvA oscillations can not be observed in torque measurements. We, therefore, use the data obtained at the lowest angle, 4$^\circ$, at which all the additional frequencies were observed. Since all the dHvA frequencies almost don't change at small angles, the frequency difference between 0$^\circ$ and 4$^\circ$ can be neglected. The dHvA frequencies in CeRhIn$_5$ at high fields do not match those under high pressure. Only the $\alpha_2$ frequency at high pressure roughly corresponds to the $N$ frequency at high field. On the other hand, the $\alpha_1$ and $\alpha_3$ frequencies at $P =$ 2.9~GPa do not match any high field frequencies.  This comparison, therefore, argues in favor of the localized $f$ electrons picture in CeRhIn$_{5}$ at high magnetic fields.

\end{document}